\documentclass[pre]{revtex4}
\usepackage[cp1251]{inputenc}
\usepackage[english,russian]{babel}

\usepackage{graphicx}
\usepackage{dcolumn}
\usepackage{eucal}
\usepackage[dvips]{epsfig}
\usepackage{amssymb}
\usepackage{amsmath}

\begin{document}


\newcommand{\bs}{\boldsymbol}
\newcommand{\mbb}{\mathbb}
\newcommand{\mcal}{\mathcal}
\newcommand{\mfr}{\mathfrak}
\newcommand{\mrm}{\mathrm}

\newcommand{\ovl}{\overline}
\newcommand{\p}{\partial}

\renewcommand{\d}{\mrm{d}}
\newcommand{\lap}{\triangle}

\newcommand{\lan}{\bigl\langle}
\newcommand{\ran}{\bigl\rangle}

\newcommand{\bse}{\begin{subequations}}
\newcommand{\ese}{\end{subequations}}

\newcommand{\be}{\begin{eqnarray}}
\newcommand{\ee}{\end{eqnarray}}

\newcommand{\ga}{\alpha}
\newcommand{\gb}{\beta}
\newcommand{\gc}{\gamma}
\newcommand{\gd}{\delta}
\newcommand{\gr}{\rho}
\newcommand{\eps}{\epsilon}
\newcommand{\veps}{\varepsilon}
\newcommand{\gs}{\sigma}
\newcommand{\gf}{\varphi}
\newcommand{\go}{\omega}
\newcommand{\gl}{\lambda}

\renewcommand{\l}{\left}
\renewcommand{\r}{\right}


\title{\bf ТОЧНЫЙ ФУНКЦИОНАЛ ПЛОТНОСТИ ДЛЯ ЭНЕРГИИ НЕРЕЛЯТИВИСТСКОЙ ЧАСТИЦЫ В ЛОКАЛЬНОМ ВНЕШНЕМ ПОЛЕ
}
\author{V.B. Bobrov $^{1}$, S.A. Trigger $^{1,2}$}
\address{$^1$ Joint\, Institute\, for\, High\, Temperatures, Russian\, Academy\,
of\, Sciences, Izhorskaia St., 13, Bd. 2. Moscow\, 125412, Russia;\\
emails:\, vic5907@mail.ru,\;satron@mail.ru\\
$^2$ Eindhoven  University of Technology, P.O. Box 513, MB 5600
Eindhoven, The Netherlands}

\begin{abstract}
На основе уравнения Шредингера получены точные выражения для энергии нерелятивистской частицы в локальном внешнем поле и потенциала внешнего поля как функционалов неоднородной плотности. На этой основе показано, что при рассмотрении более двух невзаимодействующих электронов энергия такой системы не может быть функционалом неоднородной плотности. Это означает, что лемма Хоэнберга-Кона  о том, что в основном состоянии двум различным потенциалам внешнего поля  не может соответствовать одна и та же неоднородная плотность, не может служить в общем случае обоснованием существования универсального функционала плотности. При этом положения теории функционала плотности остаются в силе при рассмотрении произвольного числа невзаимодействующих бозонов в основном состоянии из-за эффекта Бозе-коденсации.\\

 PACS number(s): 71.15.Mb, 05.30.Ch, 71.10.Ca, 52.25.Kn \\

\end{abstract}

\maketitle

Согласно лемме Хоэнберга-Кона [1,2] в основном состоянии нерелятивистской системы электронов двум различным локальным потенциалам внешнего поля  $v_1({\bf r})$ и  $v_2({\bf r})$ не может соответствовать одна и та же неоднородная плотность  $ n({\bf r})$ (кроме случая $ v_1({\bf r})- v_2({\bf r})= \mathrm{const} $). Таким образом, неоднородная плотность  $n({\bf r})$ нерелятивистской системы электронов, находящейся  в основном состоянии, однозначно определяет потенциал  $ v({\bf r})$ (с точностью до аддитивной постоянной). В случае вырождения основного состояния лемма относится к плотности  $ n({\bf r})$ любого основного состояния. В соответствии с леммой Хоэнберга-Кона принято считать, что потенциал внешнего поля  $ v({\bf r})$ является функционалом неоднородной плотности,

\begin{eqnarray}
n({\bf r})=n[v({\bf r})] \to v({\bf r})= v[n({\bf r})] + \mathrm{const}.
 \label{F1}
\end{eqnarray}
Это означает, что существует универсальное правило, по которому потенциал внешнего поля  $ v({\bf r})$ может быть найден по известной неоднородной плотности  $ n({\bf r})$, отвечающей основному состоянию рассматриваемой системы. Под этим подразумевается, что в принципе существует (хотя и не может быть явно найдено или указано) правило нахождения функции  $ v({\bf r})$   по известной функции  $ n({\bf r})$, структура которого не зависит от явного вида функций  $ v({\bf r})$ и  $ n({\bf r})$. Подчеркнем, что  этому очень сильному утверждению до настоящего времени не уделяется достаточного внимания.  При этом между леммой Хоэнберга-Кона и соотношением (1), строго говоря, нет взаимно однозначного соответствия (см.ниже).

Если же принять справедливость утверждения (1), тогда для энергии основного состояния  $E_0$ системы из  $N$ взаимодействующих электронов с гамильтонианом  $H$ во внешнем поле с потенциалом  $ v({\bf r})$, которое характеризуется волновой функцией  $\Psi_{0}$, можно записать

\begin{eqnarray}
E_0\equiv \langle\Psi_{0}|H|\Psi_{0}\rangle = E_0(N,[ \Psi_{0},v({\bf r})])=E_0[n({\bf r}),v({\bf r})],\qquad N=\int n({\bf r})d^3{\bf r}.
\label{F2}
\end{eqnarray}
Здесь учтено, что  $\langle\Psi_{0}|\Psi_{0} \rangle =1$ и   $\Psi_{0}[v({\bf r})] = \Psi_{0}[v({\bf r})+\mathrm{const}]$. В свою очередь, из (2) непосредственно следует, что величина     $F[n({\bf r})]= \langle\Psi_{0}|T+U|\Psi_{0} \rangle $, которая определяет
энергию основного состояния системы

\begin{eqnarray}
E_0[n({\bf r}),v({\bf r})]=F[n({\bf r})]+\int v({\bf r})n({\bf r})d^3{\bf r}
, \label{F3}
\end{eqnarray}
является функционалом только плотности  $n({\bf r})$ (универсальный функционал плотности).  Здесь операторы  $T$  и  $U$ – соответственно операторы кинетической энергии и энергии межчастичного взаимодействия.  Утверждение (3) является основой теории функционала плотности (DFT), широко применяемой в различных областях физики и химии (см., например, [3,4]). Однако, до настоящего времени точный вид этого универсального функционала не известен даже для невзаимодействующих между собой электронов $(U=0)$.
В связи с этим рассмотрим один нерелятивистский электрон массы  $m$ в статическом внешнем поле  $v({\bf r})$. Тогда стационарное состояние электрона, которое характеризуется некоторым набором квантовых чисел  $\alpha $, включающем спиновое квантовое число  $\sigma $, полностью определяется волновой функцией  $\Phi_{\alpha}({\bf r})$, которая удовлетворяет уравнению Шредингера

\begin{eqnarray}
\left\{-\frac{\hbar ^2}{2m}\Delta_{{\bf r}}+v({\bf r}) \right\}\Phi_{\alpha}({\bf r})= \epsilon_{\alpha} \Psi_{\alpha }({\bf r}),
\label{F4}
\end{eqnarray}
где  $\epsilon_{\alpha}$ – энергия электрона в соответствующем состоянии. Так как энергия электрона не зависит от спина, то каждое значение  $\epsilon_{\alpha}$ двукратно вырождено по спиновому квантовому числу, также как и другие значения физических величин, в том числе неоднородная плотность   $n_{\alpha}({\bf r})$. Уравнение (4) на собственные значения принято решать при выполнении граничного условия   $\Phi_{\alpha}(|{\bf r}|\to \infty)=0$ (так называемое условие на бесконечности [5]). С учетом возможности рассмотрения системы в конечном объеме  $V$ граничное условие для уравнения (4) в наиболее общем виде записывается как

\begin{eqnarray}
\Phi_{\alpha}({\bf r}\to {\bf S})=0,
 \label{F5}
\end{eqnarray}
где  $S$ – поверхность, ограничивающая объем  $V$. Учтем далее, что волновая функция  $\Phi_{\alpha}({\bf r})$ может быть рассмотрена как действительная функция [5]. Тогда

\begin{eqnarray}
n_{\alpha}({\bf r})=| \Phi_{\alpha}({\bf r})|^2= \Phi_{\alpha}^2({\bf r}),
\label{F6}
\end{eqnarray}

\begin{eqnarray}
\nabla_{\bf r}n_{\alpha}({\bf r})=2\Phi_{\alpha}({\bf r})\nabla_{\bf r}\Phi_{\alpha}({\bf r}),.\qquad \Delta_{\bf r}n_{\alpha}({\bf r})=2\Phi_{\alpha}({\bf r})\Delta_{\bf r}\Phi_{\alpha}({\bf r})+2(\nabla_{\bf r}\Phi_{\alpha}({\bf r}))(\nabla_{\bf r}\Phi_{\alpha}({\bf r})).
\label{F7}
\end{eqnarray}
Из (4)-(7) непосредственно следует, что неоднородная плотность  $n_{\alpha}({\bf r}) $ удовлетворяет уравнению на собственные значения

\begin{eqnarray}
-\frac{\hbar ^2}{4m}\Delta_{\bf r}n_{\alpha}({\bf r})+\frac{\hbar ^2}{8mn_{\alpha}({\bf r})} (\nabla_{\bf r}n_{\alpha}({\bf r}))(\nabla_{\bf r}n_{\alpha}({\bf r}))+v({\bf r})n_{\alpha}({\bf r})= \epsilon_{\alpha}n_{\alpha}({\bf r})
\label{F8}
\end{eqnarray}
с граничными условиями

\begin{eqnarray}
n_{\alpha}({\bf r\to S})=0, \qquad \nabla_{\bf r}n_{\alpha}({\bf r})|_{\bf r\to S}=0.
\label{F9}
\end{eqnarray}
Проинтегрируем теперь уравнение (8) по объему, занимаемому системой. Учтем при этом условие нормировки, которое непосредственно следует из (6),

\begin{eqnarray}
\int n_{\alpha}({\bf r})dV=1
\label{F10}
\end{eqnarray}
Тогда из (8) находим функционал плотности для энергии  $\epsilon_{\alpha} $

\begin{eqnarray}
\epsilon_{\alpha}[n_{\alpha}({\bf r}),v({\bf r})]= F^{(1)}[n_{\alpha}({\bf r})]+\int v({\bf r}) n_{\alpha}({\bf r})dV
\label{F11} ,
\end{eqnarray}

\begin{eqnarray}
F^{(1)} [n_{\alpha}({\bf r})]= F^{(1)}_0 [n_{\alpha}({\bf r})]+ F^{(1)}_W[n_{\alpha}({\bf r})],
\label{F12}
\end{eqnarray}

\begin{eqnarray}
F^{(1)}_0[n_{\alpha}({\bf r})]= -\frac{\hbar ^2}{4m}\int \Delta_{\bf r}n_{\alpha}({\bf r})dV, \qquad F^{(1)}_W[n_{\alpha}({\bf r})]=\frac{\hbar ^2}{8m}\int \frac{(\nabla_{\bf r}n_{\alpha}({\bf r}))(\nabla_{\bf r}n_{\alpha}({\bf r}))}{n_{\alpha}({\bf r})}dV.
\label{F13}
\end{eqnarray}
Здесь   $F^{(1)} [n_{\alpha}({\bf r})]$ --- универсальный функционал плотности   $F[n({\bf r})] $ (3) для одной частицы (индекс (1)). Представление функционала  $F^{(1)} [n_{\alpha}({\bf r})]$ в виде двух слагаемых в (12) обусловлено двумя причинами. Во-первых, с учетом формулы Гаусса и второго граничного условия в (9) функционал  $F^{(1)}_0[n_{\alpha}({\bf r})]$ обращается в нуль,

\begin{eqnarray}
F^{(1)}_0[n_{\alpha}({\bf r})]= -\frac{\hbar ^2}{4m}\int \Delta_{\bf r}n_{\alpha}({\bf r})dV= -\frac{\hbar ^2}{4m}\oint \nabla_{\bf r}n_{\alpha}({\bf r})d{\bf S}=0.
\label{F14}
\end{eqnarray}
Во-вторых, явный вид функционала  $F^{(1)}_W[n_{\alpha}({\bf r})]$ (13) формально в точности совпадает с так называемой the Weizsaker correction to the Thomas-Fermi kinetic energy functional [6] (см.подробнее [3,4]). Таким образом, универсальный функционал плотности  $F^{(1)}_0[n_{\alpha}({\bf r})]$ в рассматриваемом случае существует (как для основного, так и возбужденных состояний), находится точно и равен

\begin{eqnarray}
F^{(1)}_0[n_{\alpha}({\bf r})]= F^{(1)}_W[n_{\alpha}({\bf r})]
\label{F15}
\end{eqnarray}
Нетрудно убедиться прямым вычислением (см., например, [4]), что

\begin{eqnarray}
\frac{\delta F^{(1)}_W[n_{\alpha}({\bf r})]}{\delta n_{\alpha}({\bf r})} = - \frac{\hbar ^2}{4mn_{\alpha}({\bf r})}\Delta_{\bf r}n_{\alpha}({\bf r})+ \frac{\hbar ^2}{8mn_{\alpha}^2 ({\bf r})} (\nabla_{\bf r}n_{\alpha}({\bf r})) (\nabla_{\bf r}n_{\alpha}({\bf r})).
\label{F16}
\end{eqnarray}
Таким образом, уравнение (8) на собственные значения энергии  $\epsilon^{(1)}$ одной частицы во внешнем поле  $v({\bf r})$ с граничными условиями (9) является следствием вариационного уравнения для энергии   $\epsilon^{(1)} [n^{(1)}({\bf r}),v({\bf r})] $ как функционала неоднородной плотности   $n^{(1)}({\bf r})$ одной частицы в заданном внешнем поле  $v({\bf r})$,

\begin{eqnarray}
\delta \epsilon^{(1)}[n^{(1)}({\bf r})]=0
\label{F17}
\end{eqnarray}

Действительно, используя условие нормировки (10) и преобразование Лежандра, из (17) находим

\begin{eqnarray}
\frac{\delta \epsilon^{(1)} [n_{\alpha}({\bf r})]}{\delta n_{\alpha}({\bf r})} = \mathrm{const}.
\label{F18}
\end{eqnarray}

Для определения постоянной в уравнении (18) учтем, что согласно (11)-(16)

\begin{eqnarray}
\frac{\delta F^{(1)}[n_{\alpha}({\bf r})]}{\delta n_{\alpha}({\bf r})} + v({\bf r}) = \mathrm{const} = \epsilon^{(1)}.
\label{F19}
\end{eqnarray}

Таким образом, вариационное уравнение (17) эквивалентно (8) полностью аналогично соответствующему уравнению для энергии системы частиц как функционала волновой функции в квантовой механике (см., например, [5]). С учетом (16),(19) находим потенциал внешнего поля как функционал плотности, определенный с точностью до постоянной величины,

\begin{eqnarray}
v ({\bf r})+ \mathrm{const}=\frac{\delta F^{(1)}[n_{\alpha}({\bf r})]}{\delta n_{\alpha}({\bf r})} = \frac{\hbar ^2}{4mn_{\alpha}({\bf r})} \Delta_{\bf r}n_{\alpha}({\bf r})-\frac{\hbar ^2}{8mn_{\alpha}^2({\bf r})}(\nabla_{\bf r}n_{\alpha}({\bf r})) (\nabla_{\bf r}n_{\alpha}({\bf r})).
\label{F20}
\end{eqnarray}

Учитывая, что доказательство леммы Хоэнберга-Кона, как и утверждения (1) и (3) никоим образом не зависят от конкретного значения числа частиц  $N$ в рассматриваемой системе, возникает естественное желание распространить полученные результаты на случай произвольного числа невзаимодействующих электронов. Для этого, казалось бы, достаточно осуществить следующую замену:

\begin{eqnarray}
n_{\alpha}({\bf r})\to n({\bf r}),\qquad \epsilon^{(1)}\to E_0,
\label{F21}
\end{eqnarray}
вместо условия нормировки (10) использовать соответствующее условие в (2), а также принять в соответствии с (3),(11)-(13), что величина   $\langle\Psi_{0}|T|\Psi_{0} \rangle $ для системы нескольких невзаимодействующих частиц равна

\begin{eqnarray}
\langle\Psi_{0}|T|\Psi_{0} \rangle = F^{(0)}[n({\bf r})]=-\frac{\hbar ^2}{4m}\int \Delta_{\bf r}n({\bf r})dV +\frac{\hbar ^2}{8m}\int \frac{(\nabla_{\bf r}n({\bf r}))(\nabla_{\bf r}n({\bf r})) }{n({\bf r})} dV.
\label{F22}
\end{eqnarray}

Чтобы выяснить, справедливы ли соотношения (21), (22), рассмотрим систему из  $N$ невзаимодействующих электронов во внешнем поле  $v({\bf r})$. Для учета тождественности электронов такое рассмотрение наиболее удобно проводить в формализме вторичного квантования (см. подробнее [4,5]).. Тогда произвольное состояние системы из  $N$ невзаимодействующих тождественных электронов характеризуется набором так называемых «занятых» одночастичных состояний  $\alpha_1, \ldots, \alpha_N$ (см.(4)), причем в силу принципа Паули

\begin{eqnarray}
\alpha_i \neq   \alpha_j   \quad \textrm{ при }\quad i\neq j.
\label{F23}
\end{eqnarray}

Тогда энергия  $E^{(0)}$ и неоднородная плотность  $n^{(0)}({\bf r})$ в соответствующем состоянии равны [4,5]

\begin{eqnarray}
E^{(0)}(\alpha_1, \ldots, \alpha_N) = \sum_{\alpha _i}\epsilon_{\alpha_i}, \qquad n^{(0)}({\bf r}, \alpha_1, \ldots, \alpha_N)= \sum_{\alpha _i}n_{\alpha_i}({\bf r}).
\label{F24}
\end{eqnarray}

При этом  $ \epsilon_{\alpha_i } =\epsilon_{\alpha_i } [n_{\alpha}({\bf r}), v({\bf r})]$ (см.(11)-(13)) и  $ \Sigma_{\alpha_i}1=N$. Тогда из (23), (24) непосредственно следует, что энергия  $E^{(0)} $ системы из  $N(N\geq 3)$ невзаимодействующих тождественных электронов, в том числе энергия  $E^{(0)}_0 $ основного состояния,  не может являться функционалом плотности  $n^{(0)}({\bf r}) $ в заданном внешнем поле из-за нелинейности функционала  $F^{(1)}_W[n_{\alpha_i } ({\bf r})] $ (13). Очевидно, аналогичное утверждение имеет место и в отношении универсального функционала  $\langle\Psi_{0}|T|\Psi_{0} \rangle $. Это является следствием того, что потенциал внешнего поля   $v({\bf r})$ при  $N\geq 3$ не может быть представлен как функционал плотности  $n({\bf r})$ (см.(20)).

При этом в случае двух невзаимодействующих электронов, находящихся в основном состоянии, положения теории функционала плотности остаются в силе благодаря двукратному вырождению по спиновому квантовому числу  $(n({\bf r}) = 2n^{(1)}({\bf r}))$. Отметим также, что эти результаты непосредственно связаны с принципом Паули, который имеет отношение только к фермионам. В случае бозонов, которые в основном состоянии «накапливаются» на одном низшем энергетическом уровне  (Бозе-конденсация), положения  теории функционала плотности сохраняются при произвольном числе невзаимодействующих бозонов.
В результате, мы приходим к выводу, что лемма Хоэнберга-Кона [1,2], строго говоря,  не может служить обоснованием существования универсального функционала плотности в общем случае.

\section*{Acknowledgment}

This study was supported by the Netherlands Organization for Scientific Research (NWO), project no. 047.017.2006.007 and the Russian Foundation for Basic Research, project no. 07-02-01464-a.

\end{document}